\documentstyle[11pt,epsfig,psfig]{article}
\def\qed{\nobreak\kern 1em \vrule height .5em width .5em depth 0em}
\def\vbar{\mathchoice{\vrule height6.3ptdepth-.5ptwidth.8pt\kern-.8pt}
   {\vrule height6.3ptdepth-.5ptwidth.8pt\kern-.8pt}
   {\vrule height4.1ptdepth-.35ptwidth.6pt\kern-.6pt}
   {\vrule height3.1ptdepth-.25ptwidth.5pt\kern-.5pt}}

\def\date
   {\noindent Date: \today \par
    \medskip}

\newcount\subno      
\newcount\subsubno   
\newcount\appno      
\newcount\appeqno    
\newcount\tableno    
\newcount\figureno   
\def\nsection#1
   {\vskip0pt plus.1\vsize\penalty-250
    \vskip0pt plus-.1\vsize\vskip24pt plus12pt minus6pt
    \subno=0 \subsubno=0
    \addtocounter{section}{1}
\renewcommand{\thesection}{\Roman{section}}
  {\small  \noindent {\bf \thesection. #1\par}}
    \noindent}
\def\nsubsecnn#1
   {\vskip-\lastskip
    \vskip12pt plus12pt minus6pt
    {\noindent {\bf #1\par}}
    \noindent}
\begin{document}
%

\begin{center}

    {\bf Slow Coarsening in a Class of Driven Systems}

\end{center}
\begin{center}
\normalsize  Y. Kafri$^1$, D. Biron$^1$, M. R. Evans$^2$, and D. Mukamel$^1$
\\ $^1${ \it Department of Physics of Complex
Systems, The Weizmann Institute of Science, Rehovot 76100, Israel. }
\\ $^2${ \it Department of Physics and Astronomy,
University of Edinburgh, Mayfield Road, Edinburgh EH9 3JZ, United Kingdom.\\[4cm] }
\end{center}
\noindent {\bf Abstract:}
The coarsening process in a class of driven systems is studied.
These systems have previously been shown to
exhibit phase separation and slow coarsening in one dimension. 
We consider generalizations of this class of models to higher dimensions.
In particular
we study a system of three types of particles that diffuse under local conserving
dynamics in two dimensions. 
Arguments and numerical studies
are presented indicating that the coarsening process in any number of dimensions 
is logarithmically slow in time. A key feature of this behavior 
is that the interfaces separating
the various growing domains are smooth 
(well approximated by a Fermi function). This implies that the 
coarsening mechanism in one dimension is readily extendible to higher dimensions.

\date
\rule{7cm}{0.2mm}
\begin{flushleft}
\parbox[t]{12.5cm}{ }
\\[2mm]
\parbox[t]{3.5cm}{\bf PACS numbers:} 05.40.-a, 05.70.Ln, 05.70.Np, 64.60.My, 64.60.Cn      
\\[2mm]
\end{flushleft}
\normalsize
\thispagestyle{empty}
\mbox{}
\pagestyle{plain}

\newpage

\setcounter{page}{1}
\pagestyle{plain}
\setcounter{equation}{0}

\normalsize
\thispagestyle{empty}
\mbox{}
\pagestyle{plain}
\setcounter{page}{1}
\setcounter{equation}{0}
\section{Introduction}

The study of phase separation and coarsening processes has been a
subject of broad and growing interest in recent years
\cite{BRAY1}. Following a quench into an ordered phase from a
disordered one the typical linear size of domains of the ordered
phase, $\ell(t)$, grows in time. Usually, at late times the system
enters a scaling regime characterized by a single length scale
$\ell(t)$ that grows as $\ell(t) \sim t^{n}$. The value of the
exponent $n$ depends on the symmetry of the order parameter and on the
conservation laws of the system.  In two dimensions and above it has
been argued that for a scalar order parameter $n$ is either $1/3$
\cite{HUSE,LS} or $n=1/2$ \cite{AC,OJK} depending on whether or not
the dynamics conserves the order parameter, respectively. 

In one
dimension and for short-range interactions the situation is different
 due to the absence of long-range
order at any temperature $T{>}0$.
An exact solution of the non-conserving Glauber dynamics
at zero temperature gives the expected scaling $\ell(t) \sim t^{1/2}$
\cite{GLAU,BRAY2,AF}.  On the other hand, at zero temperature 
under Kawasaki dynamics where the order parameter is conserved, 
the system gets trapped in metastable states
with isolated domain walls and coarsening is arrested.
Studies of such conserving one-dimensional models in a small $T$ limit 
have shown that the growth law $\ell(t) \sim t^{1/3}$ still
persists \cite{CKS,MHS}.  However, for both conserving and
non-conserving dynamics a solution of the deterministic continuum
Ginzburg-Landau equations gives rise to a different answer.  In these
cases the domain walls interact with each other at large distance through exponentially
small forces which in the absence of diffusion leads to the growth law
$\ell(t) \sim \log(t)$\cite{NK}. 

Another situation where 
a logarithmic growth law arises is when energy barriers proportional
to the domain size have to be surmounted during the coarsening. A 
specific example of this is a three dimensional Ising model with next nearest
neighbor interactions \cite{SS}. However, in this model the slow coarsening is lost
in two dimensions.

Coarsening processes in driven systems are less well understood.
Here the dynamics
does not obey detailed balance. This gives rise to many
phenomena which do not occur in thermal equilibrium \cite{SZ,David}. For
example, several one-dimensional driven systems have been shown to
exhibit long-range order and spontaneous symmetry breaking even when
the dynamics is local \cite{SYM1,SYM2}.  A study of the coarsening
process in a one-dimensional Ising model with conserved order
parameter has shown that for a small driving field the average domain
size grows as $t^{1/2}$ in contrast to the $t^{1/3}$ behavior of the
non-driven case \cite{CBD,SKR}.  However, in the coarsening of the
two-dimensional conserved driven Ising model there is some evidence to
suggest that after defining a properly rescaled isotropic domain size
the growth law $\ell(t) \sim t^{1/3}$ is retained \cite{YRHJ}.

Recently, several one-dimensional driven systems have been shown to
exhibit coarsening and phase separation of a novel kind
\cite{EKKM1}--\cite{RSS}. The phase separation does not rely
on microscopic rates tending to zero (as is the case for equilibrium
systems in the limit of zero temperature). Instead the phase
separation is achieved through a drive which stabilizes certain domain
walls and allows ordered domains to be stable if the number of
species of domains is greater than two.
The mechanism (to be reviewed below)  leads to a slow
coarsening process in which $\ell(t) \sim \log(t)$ \cite{EKKM2}.

In this paper we consider the coarsening dynamics of an extension of
one of these models to two and higher dimensions.  We show that the slow,
logarithmic, coarsening is retained. In the ordered phase the various states
are separated by domain walls, which in one dimension are point-like objects. 
On the other hand in higher dimensions these interfaces are fluctuating 
extended surfaces. The evolution of the coarsening domains may
depend on the smoothness of these surfaces. It would thus be of interest to
explore this issue in some detail.
An analysis of the model introduced in this paper shows that the interfaces are
smooth and
have profiles described by a Fermi function with a linear potential (see
Eq. \ref{Fermi}). This is expected to be a generic feature for the class
of models under consideration.

The one dimensional version of the model we study is defined on a ring of length
$L$ where each site is occupied by one of the three types of particles
$A$, $B$ or $C$ \cite{EKKM1}.
The model evolves under a random sequential update procedure which is defined as follows:
at each time step two neighboring sites are chosen randomly, and the particles
at these sites are exchanged according to the rates
\begin{equation}
\label{eq:dynamicsy}
\begin{picture}(130,37)(0,2)
\unitlength=1.0pt
\put(36,6){$BC$}
\put(56,4) {$\longleftarrow$}
\put(62,0) {\footnotesize $1$}
\put(56,8) {$\longrightarrow$}
\put(62,13) {\footnotesize $q$}
\put(80,6){$CB$}
\put(36,28){$AB$}
\put(56,26) {$\longleftarrow$}
\put(62,22) {\footnotesize $1$}
\put(56,30) {$\longrightarrow$}
\put(62,35) {\footnotesize $q$}
\put(80,28){$BA$}
\put(36,-16){$CA$}
\put(56,-18) {$\longleftarrow$}
\put(62,-22) {\footnotesize $1$}
\put(56,-14) {$\longrightarrow$}
\put(62,-9) {\footnotesize $q$}
\put(80,-16){$AC$.}
\end{picture}
\end{equation}
\vspace{0.4cm}

\noindent The model conserves the number of particles $N_A, N_B$ and
$N_C$ of the three species.  When $q=1$ the particles undergo
symmetric diffusion and the system is disordered.  For completeness
we now review the explanation of why the system coarsens logarithmically
in time for any $q \neq 1$ \cite{EKKM2}. This argument will 
provide a starting point for the analysis in higher dimensions. 
To be explicit we consider the case $q<1$. Due to the
bias, an $A$ particle prefers to move to the left inside a $B$ domain and to
the right inside a $C$ domain. Similarly the motion of the $B$ and $C$
particles in foreign domains is biased. Consider the evolution of the
system starting from a random initial condition. The configuration is
composed of a random sequence of $A,B$ and $C$ particles. Due to the
bias a local configuration in which an $A$ domain is located to the
right of a $B$ domain is unstable, and the two domains will exchange
places on a relatively short time scale linear in the domain
size. Similarly, $AC$ and $CB$ domain walls are unstable. In contrast
domain walls of the type $AB,BC$ and $CA$ are long lived. Thus, after
a short time the system will rearrange into a state of the type
$\ldots AABBBCCAAABBCCC \ldots$ in which only stable domain walls are
present.  The evolution of this state will proceed by slow diffusion
against a bias in which, for example, an $A$ particle crosses an
adjacent $C$ domain. The time scale for such a process to occur is
$q^{-\ell}$ where $\ell$ is the typical domain size in the system. This
suggests that the average domain size in the system grows as $\ln t /
| \ln q |$. Eventually the system will phase separate into three domains
of the three species of the form $A \ldots AB \ldots BC \ldots C$ with
small density fluctuations around the domain walls, leaving the bulk
of the domains pure. In another model in
the class this has been referred to as strong phase separation
\cite{LR2}.
In the thermodynamic limit these domains do not remix, implying
breaking of translational invariance. For the specific case
where the number of particles of each species is equal it was shown
that the local dynamics obeys detailed balance with respect to a
long-range Hamiltonian. Indeed, using this Hamiltonian it was
rigorously proved that the model is completely phase separated into
pure domains in the steady state \cite{EKKM1,EKKM2}. 

The model we study is the simplest generalization of the above
one-dimensional model to higher dimensionality. To be explicit, we
consider the model in $d=2$ dimensions but the results presented
in this paper are also valid in higher dimensions.
We introduce a second
lattice direction in which the particles perform unbiased
diffusion. That is we study an $L_x \times L_y$ lattice, with periodic
boundary conditions, in which at each time step two neighboring sites
are chosen at random. If the two sites lie along the $x$ axis the
particles perform a biased diffusion defined by
Eq. \ref{eq:dynamicsy} while if they lie in the $y$ axis the
particles perform an unbiased diffusion. Thus, along the $y$ axis the
dynamics is defined through the rates
\begin{equation}
\label{eq:dynamicsx}
\begin{picture}(130,37)(0,2)
\unitlength=1.0pt
\put(36,6){$BC$}
\put(56,4) {$\longleftarrow$}
\put(62,0) {\footnotesize $1$}
\put(56,8) {$\longrightarrow$}
\put(62,13) {\footnotesize $1$}
\put(80,6){$CB$}
\put(36,28){$AB$}
\put(56,26) {$\longleftarrow$}
\put(62,22) {\footnotesize $1$}
\put(56,30) {$\longrightarrow$}
\put(62,35) {\footnotesize $1$}
\put(80,28){$BA$}
\put(36,-16){$CA$}
\put(56,-18) {$\longleftarrow$}
\put(62,-22) {\footnotesize $1$}
\put(56,-14) {$\longrightarrow$}
\put(62,-9) {\footnotesize $1$}
\put(80,-16){$AC$.}
\end{picture}
\end{equation}
\vspace{0.4cm}

Here for simplicity the transition rates are taken to be $1$,
although choosing a different rate for hops in the $y$ 
direction would not affect the results obtained in this paper.
We expect other generalizations of the model, where
the motion in the $y$ direction is biased in a manner
similar to that in the  $x$ direction, to have the same generic
behavior. We return to this point in the discussion.

The model conserves the total number of particles of each species. As
will be demonstrated below,
in two dimensions the rates do not satisfy detailed
balance for $q \ne 1$ irrespective of the number of particles of
each species. Thus the system is generically far from thermal equilibrium.

In the following we show, by numerical and analytical methods, that
the slow logarithmic coarsening persists in the two dimensional model
and also, we argue, in higher dimensions.  In two dimensions the
system forms on short time scales stripes of $A,B$ and $C$ particles
which are aligned along the $y$ direction and are ordered along the
$x$ direction in the form $\ldots AABBBCCAAABBCCC \ldots$.  Thus the
interfaces between three phases are lines whose fluctuations must be
considered. If the interfaces are smooth the argument suggesting a
logarithmic coarsening in the one-dimensional model should also apply
in the two-dimensional model. That is, the average tunneling time of a
domain would be exponentially small in the average domain size. However,
if the interfaces are rough such that they can come close to
each other it might be that different behavior would prevail.  Thus,
an important question is: how rough is the interface?  What we find
is that the interfaces are indeed smooth and the typical size of a
stripe grows as $\ell(t) \sim \log(t)$.

The paper is organized as follows: it is first shown in Section 2 that
detailed balance cannot be satisfied in two dimensions (in contrast to
the one-dimensional case). In Section 3 the coarsening of the model is
studied numerically and a simple interface model which captures the
essential physics of an interface between two phases is formulated.
The model is solved and shown to have a smooth interface.  In Section
4 we discuss the generality of the results both to other two
dimensional models and to higher dimensions.

\section{Lack of detailed balance in the model}

To argue that the model presented above does not satisfy detailed balance
(except for $q=1$) we first
review the situation in the one-dimensional version of the model.
In this case it was shown \cite{EKKM2} that when the number
of particles of each species is equal, that is when $N_A=N_B=N_C$, detailed balance is
satisfied by the rates given in Eq. \ref{eq:dynamicsy}
for arbitrary $q$. However,
when the number of particles of two species are not equal, say $N_A \neq N_B$, detailed
balance is not satisfied. To see this consider
an arbitrary set of configurations $1,2 \ldots,k$.
Let $W(i \rightarrow j)$ be the transition rate from configuration $i$ to $j$. A
necessary and sufficient condition for the existence of detailed balance \cite{David}
is that for any given set of $k$ states the following equality is satisfied:
\begin{equation}
W(1 \rightarrow 2)W(2 \rightarrow 3) \ldots W(k \rightarrow 1) = W(1 \rightarrow k)
W(k \rightarrow k-1) \ldots W(2 \rightarrow 1)\;.
\label{dbcondition}
\end{equation}
That is for any closed loop in configuration space the product of rates
going along one direction should be equal to the product of rates going in
the opposite direction. 

We now apply this criterion to show that in $d=1$ detailed balance 
is not satisfied when the three densities are unequal. Consider
for simplicity the fully phase separated state
$A \ldots AB \ldots BC \ldots C$. Take the rightmost $A$ particle
and move it to the right until it has traversed both the $B$ and
the $C$ domains. The resulting configuration is a fully phase
separated state translated by one lattice unit with respect to the
starting configuration. Repeating this process $N$ times one
returns to the starting configuration. The product of the microscopic
rates involved in this process is $q^{N N_B}$. Carrying out a similar
process but in the reverse direction leads again to the starting
configuration. For this path the product of rates is $q^{N N_C}$.
Thus according to the criterion (\ref{dbcondition}) detailed balance
is not satisfied when $N_B \ne N_C$, or when any two of the densities
are not equal. It is not difficult to show that when $N_A=N_B=N_C$
detailed balance is satisfied \cite{EKKM1,EKKM2}.

Next, consider the two-dimensional model. In this case whatever the number
of particles of each species one can always choose as 
a starting configuration a state where on one
of the rows
the numbers of particles of each species are not equal. Repeating the argument
above, given for the one dimensional model, by considering
particle exchanges along this row we immediately see that due
to the unequal number of particles detailed balance
does not hold.

Therefore, in contrast to the one-dimensional model, 
whatever that number of particles of each species detailed balance is never
satisfied in two dimensions. It is straightforward to see that this argument
also implies lack of detailed balance in higher dimensional generalizations of the model.

\section{Coarsening in two dimensions}

To study the dynamics of the model in two dimensions we first
show numerically that on a short time scale, the system indeed
evolves into a striped state composed of sequences of $A,B$ and $C$ 
domains aligned in the $x$ direction. We then study the slow coarsening 
process of these stripes and argue that the typical length scale
increases logarithmically with time.

\subsection{Monte Carlo simulation}

\begin{figure}
\epsfxsize 16 cm
\epsfysize 16 cm
\hspace{0cm}\epsfbox{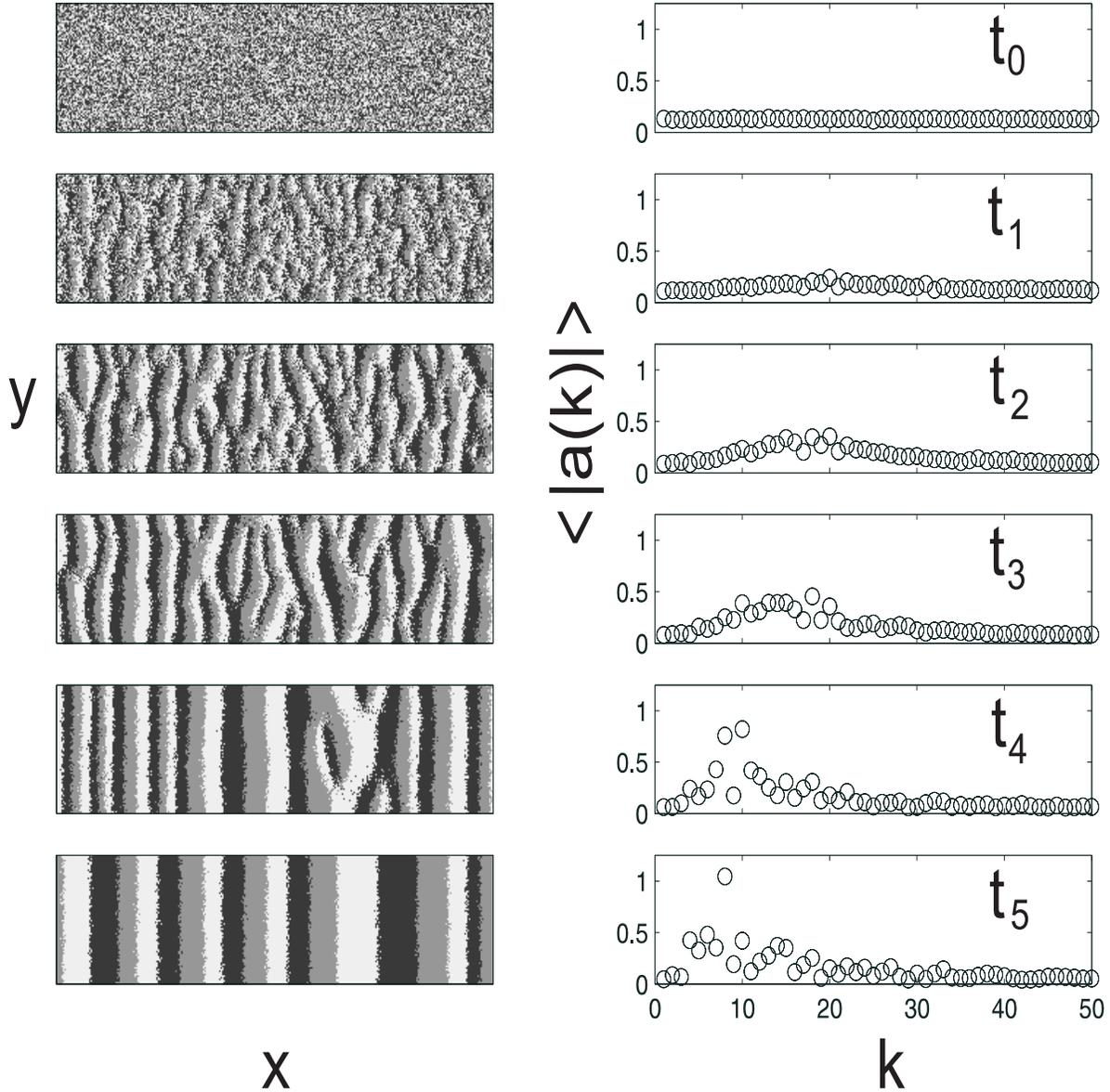}
\caption{On the left hand side one can see from top to bottom
the evolution of a typical system
for $t_0=0,t_1=30,t_2=66,t_3=146,t_4=920,t_5=383000$ Monte Carlo sweeps when $q=0.15$.
The different species of particles are represented as different gray scales.
On the right hand side the first 50 Fourier components of
$\langle \vert {\rm a}(k) \vert \rangle $
averaged over a hundred simulations for the same times are presented and the same value of $q$.
Similar results were obtained for different values of $q$.}
\label{evol}
\end{figure}

In order to study the short time behavior of the model and
demonstrate the flow into a striped state,
Monte Carlo simulations were performed for various values of $L_x$ and $L_y$.
Starting from a random initial condition we studied the case with equal densities
of particles of each species. A typical evolution of a system is presented
in Fig. \ref{evol} for a lattice with $L_x=300$
and $L_y=80$ with $q=0.15$. 
One can clearly see that on short time
scales a striped structure evolves in which stripes along
the $y$ direction develop. 
As expected, the stripes
are ordered in sequence $A,B,C$ along the $x$
direction. On short time scales, when the distance
between neighboring domain walls is comparable to their width,
the stripes are
strongly fluctuating. At this stage the evolution of the structure
through the formation of topological defects is relatively fast.
As time progresses the asymptotic regime where the stripes
are smooth is reached. In this regime the dynamics is dominated 
by the slow coarsening mechanism described in this paper.
Due to the slow time scales involved in the simulations
of the coarsening process our data does not allow us to demonstrate
quantitatively that the coarsening is logarithmic in time. However in the
next section arguments
will be presented which support this growth law.

To analyze the data we have calculated
the Fourier transform of the density profile along the $x$ direction for each value of $y$:
\begin{equation}
{\rm a}_y(k)=\sum_{x=0}^{L_x-1} \tilde{{\rm a}}_y(x) \exp{\left( \frac{-2 \pi i kx}{L_x} \right)}\;.
\label{ft}
\end{equation}
Here $\tilde{{\rm a}}_y(x)$ is
equal to $-1,0$ or $1$ if the site $(x,y)$ is occupied by an $A,B$ or $C$ particle
respectively. It thus corresponds to the density difference of the $A$ and $C$ species.
We then calculated $\langle \vert {\rm a}(k) \vert \rangle $, the average of $| {\rm a}_y(k) |$
for all values of $y$ and over one hundred simulations. The results are also presented
in Fig. \ref{evol}. One can
clearly see that stripes of a characteristic size form.
This is accompanied by the emergence of a typical
Fourier mode. As time progresses the $k$ value of the typical Fourier
mode decreases as expected in a coarsening system.

\begin{figure}
\epsfxsize 12 cm
\epsfysize 12 cm
\hspace{2.5cm}\epsfbox{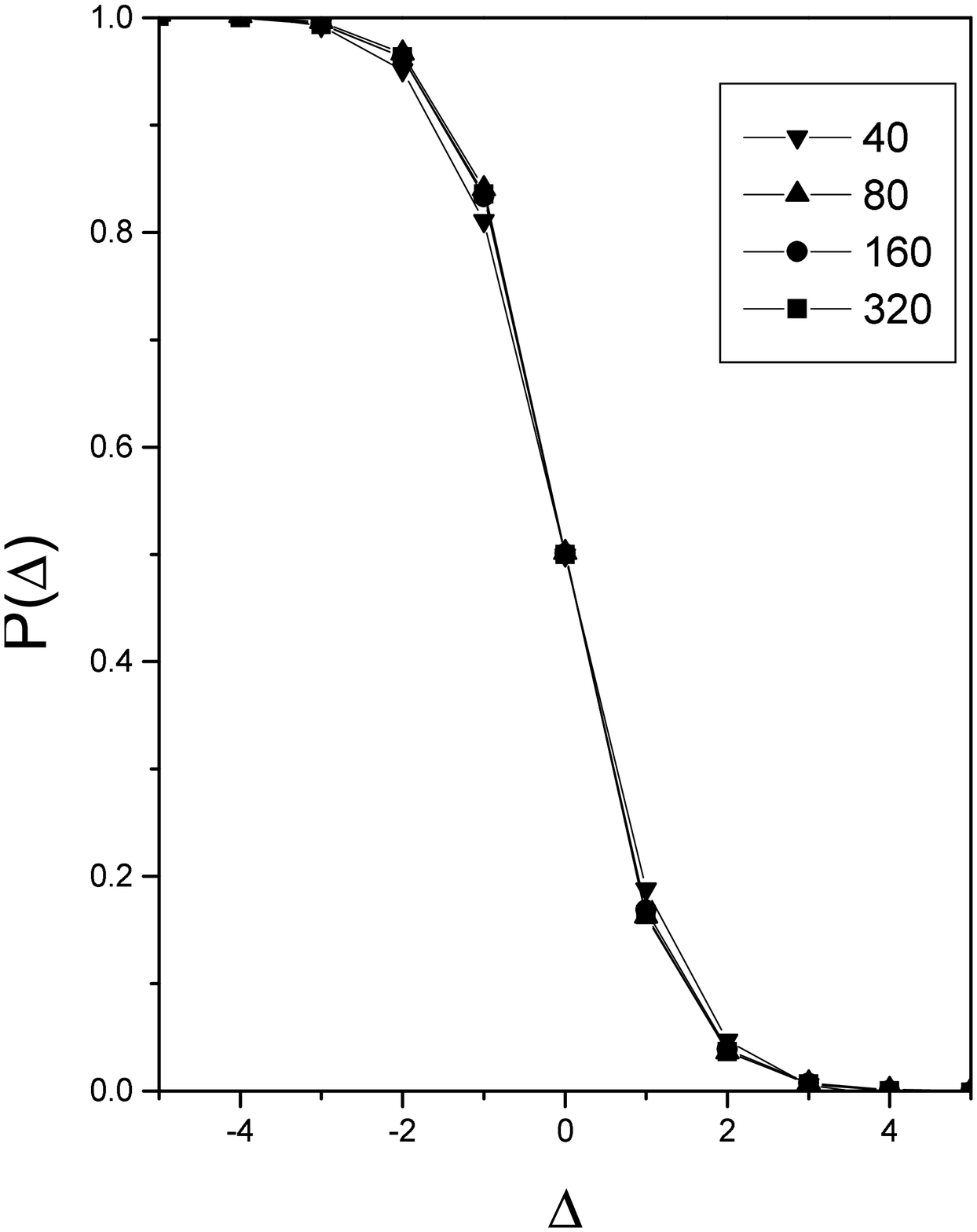}
\caption{Profiles obtained from Monte Carlo simulations
for $L_x=300$ and $L_y=40,80,160,320$ and $q=0.15$.
$\Delta$ is the distance from the center of the profile. 
The simulation results for each $L_y$ value is averaged over $40$
profiles. Similar results were obtained for different values of $q$.}
\label{profile1}
\end{figure}

It is of interest to examine the considerations presented above
for the existence of phase separation and slow coarsening
in one dimension and to check whether
the argument can be extended to two and higher dimensions.
In $d$ dimensions the $A,B,C$ domains are separated by
fluctuating $d{-}1$ dimensional surfaces. For the argument
for slow coarsening to go through unmodified, we require
that these surfaces are smooth, that is of finite extent in
the growth direction. Otherwise, if they were rough, the
fluctuating distance between neighboring surfaces might
significantly affect the coarsening process.
Therefore, we have studied numerically the smoothness of the 
interfaces in this model. To do this we
consider the density of, say, $A$ particles near the $AB$ domain wall. 
In Fig. \ref{profile1} such profiles obtained
from Monte Carlo simulations of the three species model are presented.
The simulations were performed for a given $q$ at fixed value of $L_x$
and varying $L_y$. They were obtained by extracting the density profile
of a given species of particles near a domain wall and averaging over
$40$ sets of data. One can see that the interface is smooth since
the density profile reaches a steady-state form of finite extent in the
$x$ direction, as $L_y$ is increased.

We have thus established numerically that the model evolves towards 
a striped state and that the interfaces between the phases are smooth. 
Next, an
interface model which captures the essential statistical properties
of an interface is studied
and shown to predict a smooth interface whose profile agrees very well with the
numerics.

\subsection{Interface model}

To construct a simple model that well approximates the
behavior of the actual interfaces we begin by assuming
that the system is already
in a striped configuration in which interfaces are well separated. The
density of, say, the $C$ particles near an $AB$ interface is then
negligibly small.  This assumption will be shown to be consistent with
a smooth interface.  Also any current of particles is exponentially
small in the domain size and can be neglected, implying that the
average velocity of the interface vanishes.
We may thus study the interface properties
by considering only $A$ and $B$ particles placed on a lattice with closed
(reflecting) boundary conditions in the direction of the bias.  As we will
show such a model reduces to a model of particles with excluded volume
interaction placed in a gravitational-like field in the $x$ direction.
For such systems the
interface is smooth for any $q \neq 1$.  This allows us to derive an
exact form for the density profile near an interface which is valid
for all dimensions. As will be argued later this result applies to a
wide variety of models considered previously and is not specific to
the model studied here. For example a similar profile is obtained
in \cite{LR2} for a one-dimensional model.

We define the interface model on an $l_x \times l_y$ lattice with
periodic boundary conditions in the $y$ direction and closed boundary
conditions in the $x$ direction. That is, the model is defined on a
lattice with closed boundary condition in the direction of the bias.
We choose the densities such that half the lattice sites of the lattice
are occupied by $A$ particles and the others are occupied by $B$ particle.
Therefore the center (average $x$ position)
of the $AB$ interface is at $l_x/2$. The dynamics of the
model is identical to that of the three species model (see
Eqs. \ref{eq:dynamicsy} and \ref{eq:dynamicsx}). However, all dynamics in
which $C$ particles would participate are irrelevant as there are no such
particles present.

In this model an $A$ can be thought of as a particle while a $B$
as a vacancy. Within this picture we have particles, placed on a lattice,
which are biased to move in a preferred direction along $x$.
Due to the closed boundary conditions, which imply a
zero current, it is easy to show
that the model satisfies detailed balance with respect to the steady
state weight (unnormalized probability)
\begin{equation}
W({\cal C})= q^{\sum_{x,y} x A_{x,y}},
\label{weight}
\end{equation}
where $A_{x,y}=1 \; (0)$ if, the site ($x,y$) in configuration ${\cal C}$ is
occupied (empty), and the sum runs over all lattice sites.
The steady state weight describes
particles placed in a linear potential
with an excluded volume interaction. Note that the above expression
is valid whatever the rates for exchange $A$ and $B$ in the $y$
direction.

The solution of the density profile is obtained in straightforward
analogy to a Fermi gas. Since
each lattice site ($x,y$) can be occupied by only a single particle it can
be considered as a state of a Fermion with energy $\epsilon_{x,y}=x$. The 
mean number of particles at site ($x,y$), $n_{x,y}$,
is given by the Fermi distribution with a temperature $T=1/ \vert \ln q \vert \;$:
\begin{equation}
n_{x,y}=\frac{1}{q^{-(x-l_x/2)}+1}\;.
\label{Fermi}
\end{equation}
As expected the distribution is independent of $y$. Here $l_x/2$ plays the role of a chemical
potential marking the center of the profile. It is easy to check
the condition that the average density in indeed $0.5$, namely
$\sum_{x,y} n_{x,y}= l_x l_y/2$. To see that this is satisfied
for any value of $q$ one uses the relation
\begin{equation}
n_{l_x/2+\Delta,y}+n_{l_x/2-\Delta,y}=1
\end{equation}
where $\displaystyle \Delta=x-l_x/2$. The density profile
of the interface is given by $P(x)=n_{x,y}$.
It has a finite width, $1/|\ln q |$, as expected for a 
smooth interface.

\begin{figure}
\epsfxsize 12 cm
\epsfysize 12 cm
\hspace{2.5cm}\epsfbox{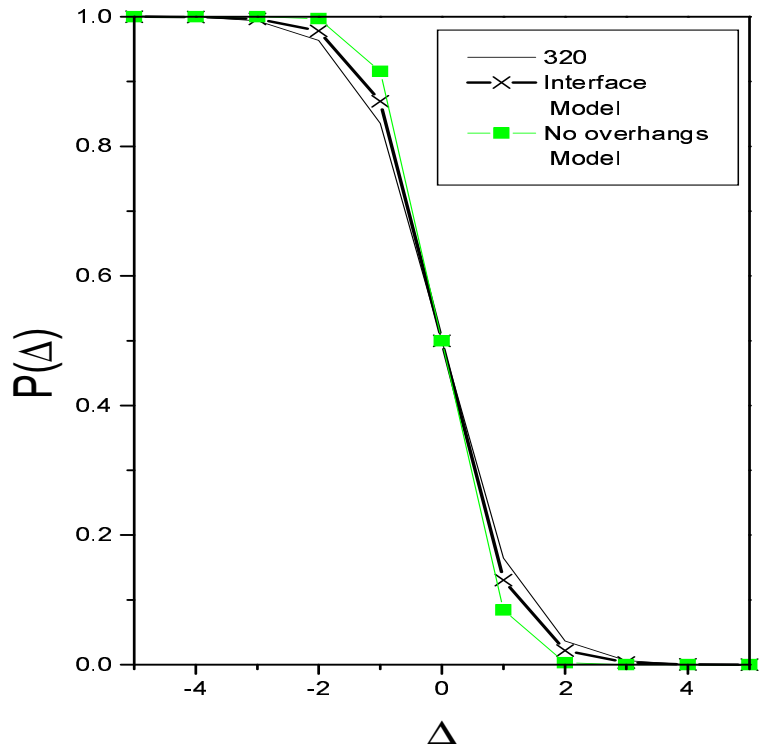}
\caption{A comparison of the profiles obtained from the interface model
and the toy model with no overhangs with the profile obtained from Monte Carlo
simulations for $L_x=300$ and $L_y=320$. 
$\Delta=x-l_x/2$ is the distance from the center of the profile.}
\label{exact}
\end{figure}

A comparison of $P(x)$ of the interface model and the profile of the Monte Carlo
simulations is given in Fig. \ref{exact}. One can see that the interface model
agrees very well with the simulations. 

We have thus established that the interfaces in the model
are smooth. This in turn implies that the argument which was presented 
for the one-dimensional version of the model is valid and predicts
a logarithmic growth law for the domain size.

The interface studied above includes configurations with bubbles
and overhangs. In the study of interfaces it is often assumed that
bubbles and overhangs may be neglected without affecting the macroscopic
properties, so that the interface may be described by a single valued
function. In the present context  we can study the validity
of such an approximation since  the profile both with
(see Eq. \ref{Fermi}) and without (see below) bubbles and
overhangs can be computed.
Thus we consider only those configurations in which the particles
fill the left part of the system, up to a `height' $h_y$ which is
the number of particles in row $y$.

By performing the sum in Eq. \ref{weight} for
each value of $y$ one obtains the following weight function:
\begin{equation}
W({\cal C})= q^{\frac{1}{2} \sum_y h_y^2},
\label{toyweight}
\end{equation}
where a term of the form $\sum_y h_y$ has been neglected, as it is
a constant equal to the total number of particles in the system.

The properties of the interface can easily be obtained by working in
the grand canonical ensemble. We thus introduce a chemical potential
to respect the constraint that the total height of the system $\sum_y h_y
=l_x l_y /2$. The weight of a configuration in this ensemble
is thus
\begin{equation}
W_{gc}({\cal C})= q^{\frac{1}{2} \sum_y (h_y-l_x/2)^2}.
\label{toyweightgc}
\end{equation}
The profile of the density along the $x$ direction is then given by
\begin{equation}
P(x)=\sqrt{\frac{2 \vert \ln q \vert}{\pi}} \int_x^{\infty}dh q^{\frac{1}{2} (h-l_x/2)^2},
\end{equation}
where the continuum limit is assumed and the upper limit of the integral was
taken to be infinity. This gives
\begin{equation}
P(x)=\frac{1}{2} {\rm erfc}\Big( (x-l_x/2)\sqrt{| \ln q |/2}\Big),
\label{prof}
\end{equation}
where ${\rm erfc}$ is the complementary error function. As expected the model
with no overhangs also predicts a smooth interface.

In Fig. \ref{exact} the resulting profile is compared with the
one obtained by the numerical simulations. It is evident that
although the two profiles agree in general, the fit is not as good
as that with the Fermi function profile. This is to be expected since,
for example, in the regions where the density of particles is low,
the no overhangs assumption becomes a less accurate approximation.
In these regions excluded volume does not play an important role.
Due to the linear potential the profile decays as
$P(x) \simeq q^{(x-l_x/2)} \propto \exp(- \alpha x)$ with $\alpha$ a constant.
However neglecting overhangs gives rise
to a decay of the form $P(x) \propto \exp(-\beta x^2)$ with $\beta$ a constant.

\section{Discussion}
In this work we have studied a generalization to two dimensions
of the $ABC$ model for phase separation \cite{EKKM1,EKKM2}. We have shown
that the slow logarithmic coarsening of the one-dimensional case persists.
Further, we have shown that the interface between the slowly coarsening domains
is smooth and can be described by a Fermi function with a linear potential 
(see Eq. \ref{Fermi}).

We next turn to the question of the generality of our results. First note that
in the generalizations of the $ABC$ model to more than three species \cite{EKKM2},
and in the models introduced by Lahiri
et al \cite{LR,LR2} and by Arndt et al \cite{AR1,AR2}
the mechanism which leads
to coarsening in the system is the same as in the $ABC$ model. 
Namely, the 
presence of stable domain boundaries is due to a bias which drives particles
of one species towards their own domain with a constant velocity. In the model
of Lahiri et al which comprises two coupled rings each containing two species
one has four different types of domain. The dynamics is then governed by
four stable interfaces separating these domains, in analogy with the model
considered in the present work.
In the model of Arndt et al there are three types of domain separated by three 
interfaces. (One of the stable domain walls corresponds to $q>0$ in Eq. \ref{Fermi} and
the other 
two correspond to $q=0$, i.e. a zero temperature Fermi function.)
Therefore, we expect that the analysis of the present paper is applicable to extensions
of these models to higher dimensions. This would imply that coarsening processes
in these models are logarithmically slow.

\begin{figure}
\epsfxsize 14 cm
\epsfysize 14 cm
\hspace{0cm}\epsfbox{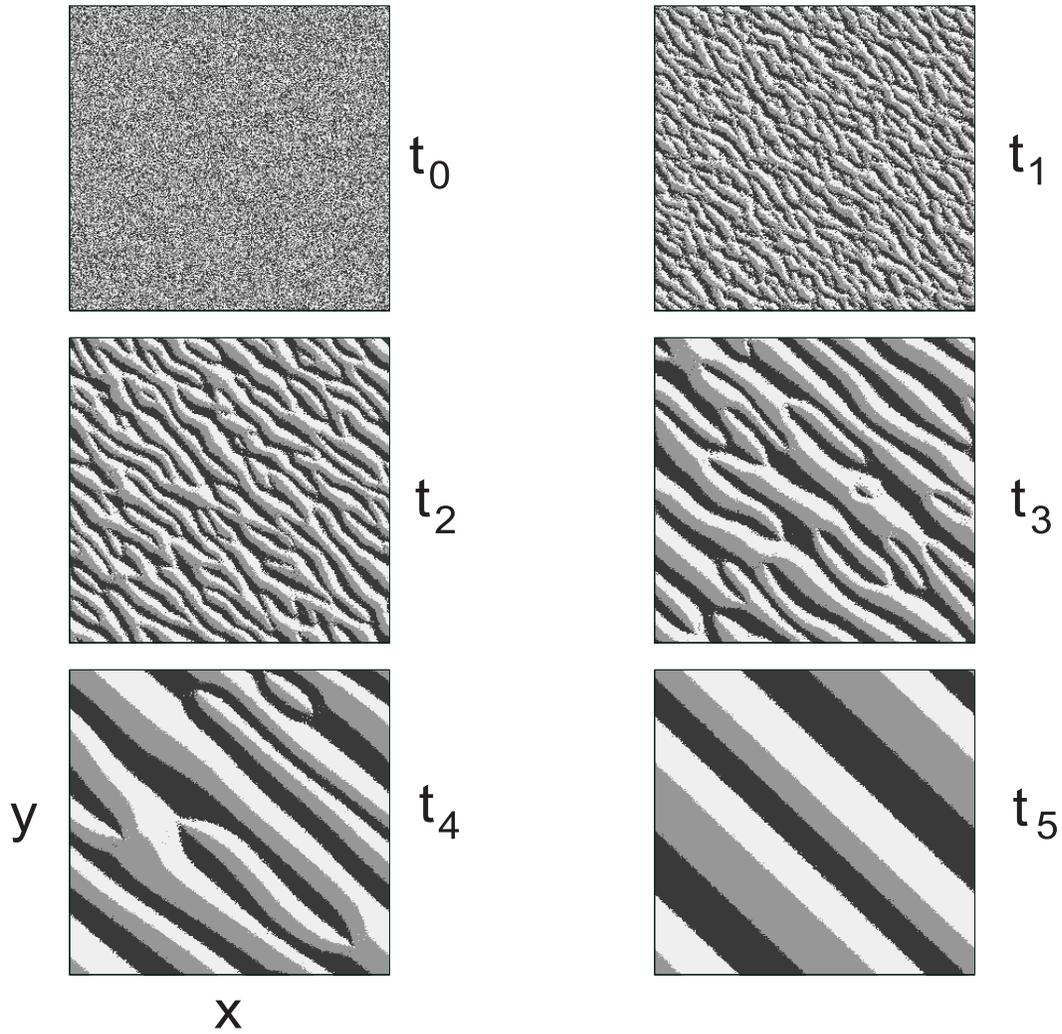}
\caption{Characteristic configurations during the evolution of a system with $q=r=0.15$
for $t_0=0,t_1=30,t_2=66,t_3=146,t_4=4440,t_5=383000$ Monte-carlo sweeps 
and $L_x=L_y=300$.
The different species of particles are represented as different gray scales.
Similar results were obtained for different values of $q$.}
\label{diag}
\end{figure}

The generalization of the model considered so far is highly anisotropic in the 
sense that the dynamics along the $x$ and $y$ direction are qualitatively different:
while the diffusion is biased in one direction it is symmetric on the other. It is 
of interest to examine the case where the diffusion along both axes (or all
axes in higher dimensions) has some bias.
Let us consider such a generalization of the $ABC$ model to two dimensions where
the exchanges along the $y$ direction have rates 
\begin{equation}
\label{eq:dynamicsxnew}
\begin{picture}(130,37)(0,2)
\unitlength=1.0pt
\put(36,6){$BC$}
\put(56,4) {$\longleftarrow$}
\put(62,0) {\footnotesize $1$}
\put(56,8) {$\longrightarrow$}
\put(62,13) {\footnotesize $r$}
\put(80,6){$CB$}
\put(36,28){$AB$}
\put(56,26) {$\longleftarrow$}
\put(62,22) {\footnotesize $1$}
\put(56,30) {$\longrightarrow$}
\put(62,35) {\footnotesize $r$}
\put(80,28){$BA$}
\put(36,-16){$CA$}
\put(56,-18) {$\longleftarrow$}
\put(62,-22) {\footnotesize $1$}
\put(56,-14) {$\longrightarrow$}
\put(62,-9) {\footnotesize $r$}
\put(80,-16){$AC$.}
\end{picture}
\end{equation}
\vspace{0.4cm}

\noindent while the exchanges along the $x$ axis are as in Eq. \ref{eq:dynamicsy}.
On some coarse grained scale (where lattice effects may be ignored)
the dynamics is effectively uniaxial. The preferred direction  
in the $x$--$y$ plane for exchange of particles is at an angle
$\theta=\arctan \left( \frac{1-q}{1-r} \right)$ with respect to the $y$ direction.
The effective exchanges
in the normal  direction are symmetric. The effect of this
would be for stripes to form perpendicular to the preferred direction. At large
scales the model reduces to the model which we have studied in this paper. Similarly,
generalizations of the $ABC$ model in higher dimensions are effectively uniaxial and 
domains form with interfaces which are $d{-}1$
dimensional hyperplanes orthogonal to the preferred direction.

To test this scenario we carried out Monte Carlo simulations of the model in 
two dimensions with $r{=}q$. 
In Fig. \ref{diag} the coarsening process is illustrated. One sees that after a transient
time stripes oriented perpendicular to the $(1,1)$ direction are formed confirming the above
picture.

Studies of phase separation and coarsening in driven systems such as that
of the present paper and the works reviewed in the introduction, have suggested that
the processes
involved are rather diverse. They range from relatively fast power law 
to slow logarithmic coarsening. For further examples of coarsening in related systems
see \cite{SHZ,KSZ,FKB}. It would be very interesting to construct a more
general framework within which all these coarsening processes 
could be understood and classified.

\noindent {\bf Acknowledgments:} We thank Rashmi Desai for interesting discussions.
MRE thanks the Einstein Center for support and DM acknowledges an EPSRC Visiting
Fellowship. Support of the Israeli Science Foundation is gratefully acknowledged.


\begin{thebibliography}{99}
\bibitem{BRAY1}
For a review see A. J. Bray, {\it Adv. Phys.} {\bf 43}, 357 (1994).

\bibitem{HUSE}
D. A. Huse, {\it Phys. Rev. B} {\bf 34}, 7845 (1986).

\bibitem{LS}
I. M. Lifshitz and V. V. Slyozov, {\it J. Chem. Solids} {\bf 19}, 35 (1961).

\bibitem{AC}
S. M. Allen and J. W. Cahn, {\it Acta. Metall.} {\bf 27}, 1085 (1979).

\bibitem{OJK}
T. Ohta, D. Jasnow and K. Kawasaki, {\it Phys. Rev. Lett.} {\bf 49}, 1223 (1982).

\bibitem{GLAU}
R. J. Glauber, {\it J. Math. Phys.} {\bf 4}, 294 (1963).

\bibitem{BRAY2}
A. J. Bray, {\it J. Phys. A} {\bf 22}, L67 (1990).

\bibitem{AF}
J. G. Amar and F. Family, {\it Phys. Rev. A} {\bf 41}, 3258 (1990).

\bibitem{CKS}
S. J. Cornell, K. Kaski and R. B. Stinchcombe, {\it Phys. Rev. B} {\bf 44}, 12263
(1991).

\bibitem{MHS}
S. N. Majumdar, D. A. Huse and B. D. Lubachevsky, {\it Phys. Rev. Lett.} {\bf 73},
182 (1994).

\bibitem{NK}
T. Nagai and K. Kawasaki, {\it Physica A} {\bf 134}, 483 (1986).

\bibitem{SS}
J. D. Shore, M. Holzer and J. P. Sethna, {\it Phys. Rev. B} {\bf 46} 11376 (1992).

\bibitem{SZ}
For a review see B. Schmittmann and R. K. P. Zia, {\it Statistical
Mechanics of Driven Diffusive Systems}, edited by C. Domb and J. L. Lebowitz,
 Phase Transitions and Critical Phenomena Vol. 17 (Academic, London, 1995).

\bibitem{David}
D. Mukamel in {\it Soft and Fragile Matter: Nonequilibrium Dynamics,
Metastability and Flow}, Eds. M. E. Cates and M. R. Evans
(Institute of Physics Publishing, Bristol, 2000); condmat/0003424.

\bibitem{SYM1}
M. R. Evans, D. P. Foster, C. Godr\`eche and D. Mukamel, {\it Phys. Rev. Lett.} {\bf 74},
 208 (1995); {\it J. Stat. Phys.} {\bf 80}, 69 (1995).
 
\bibitem{SYM2}
C. Godr\`eche, J-M. Luck, M. R. Evans, D. Mukamel, S. Sandow and E. R. Speer,
{\it J. Phys. A} {\bf 28}, 6039 (1995).

\bibitem{CBD}
S. J. Cornell and A. J. Bray, {\it Phys. Rev. E} {\bf 54}, 1153 (1996).

\bibitem{SKR}
V. Spirin, P. L. Krapivsky and S. Redner, {\it Phys. Rev. E} {\bf 60}, 2670 (1999). 

\bibitem{YRHJ}
C. Yeung, T. Rogers, A. Hernandez-Machado and D. Jasnow, {\it J. Stat. Phys.} {\bf 66},
 1071 (1992).

\bibitem{EKKM1}
M. R. Evans, Y. Kafri, H. M. Koduvely and D. Mukamel, {\it Phys. Rev. Lett.} {\bf 80},
425 (1998).

\bibitem{EKKM2}
M. R. Evans, Y. Kafri, H. M. Koduvely and D. Mukamel, {\it Phys. Rev. E} {\bf 58}, 
2764 (1998).

\bibitem{LR}
R. Lahiri and S. Ramaswamy, {\it Phys. Rev. Lett.} {\bf 79}, 1150 
(1997).

\bibitem{LR2}
R. Lahiri, M. Barma and S. Ramaswamy, {\it Phys. Rev. E}
{\bf 61}, 1648 (2000).
 
\bibitem{AR1}
P. F. Arndt, T. Heinzel and V. Rittenberg, {\it J. Phys. A} {\bf 31}, L45 (1998)

\bibitem{AR2}
P. F. Arndt, T. Heinzel and V. Rittenberg, {\it J. Stat. Phys.} {\bf 97}, 1 (1999).

\bibitem{RSS}
N. Rajewsky, T. Sasamoto and E. R. Speer, condmat/9911322.


\bibitem{SHZ}
B. Schmittmann, K. Hwang and R. K. P. Zia, {\it Europhys. Lett.} {\bf 19} 19 (1992).

\bibitem{KSZ}
G. Korniss, B. Schmittmann and R. K. P. Zia, {\it Europhys. Lett.} {\bf 32} 49 (1995); 
{\it Europhys. Lett.} {\bf 45} 431 (1999).

\bibitem{FKB}
L. Frachebourg, P. L. Krapivsky, and E. Ben-Naim, {\it Phys. Rev. E} {\bf 54} 6186 (1996).


\end{thebibliography}
\end{document}